\documentclass[11pt]{article}
\usepackage{amsmath}
\usepackage{amsfonts}   
\usepackage{amssymb}
\begin{document}
\date{}
\title{\textbf{Two Dimensional Noncommutativity and Gravitational Quantum Well}}
\author{{Saurav Samanta}\thanks{E-mail: saurav@bose.res.in}\\
\\\textit{S.~N.~Bose National Centre for Basic Sciences,}
\\\textit{JD Block, Sector III, Salt Lake, Kolkata-700098, India}}
\maketitle
                                                                                
\begin{quotation}
\noindent \normalsize 
\end{quotation}
Nesvizhevsky {\it et al.}\cite{nes} performed an experiment to study the problem of gravitational quantum well. The ``well" in which a quantum particle (neutron was used in the experiment) bounces back and forth, was formed by placing a horizontal reflecting mirror in the Earth's gravitational field. The experimental results were in reasonable agreement with the theoretical results. So if there is any noncommutative (NC) effect that must be within the error bars. In this article, we set this problem on a noncommutative phase space to calculate the upper bounds of the noncommutative parameters. A detailed description of this method is given in our paper \cite{saurav}.

We consider a two dimensional NC plane where the coordinates ($y$) and the momenta ($q$) satisfy the algebra
\begin{eqnarray}
[y_i,y_j]=i\theta\epsilon_{ij}, \  \  \ [q_i,q_j]=i\eta\epsilon_{ij}, \  \  \ [y_i,q_j]=i\hbar\delta_{ij}.
\label{4}
\end{eqnarray}
This type of NC algebra had already appeared in the context of generalized Landau problem \cite{rab}. In order to find a differential representation of the phase space variables, we consider a general representation of the form
\begin{eqnarray}
y_i\rightarrow y_i, \  \  \ q_i\rightarrow -ia\hbar\frac{\partial}{\partial y_i}+b\hbar\epsilon_{ij}\frac{\partial}{\partial y_j}+c\frac{\eta}{\hbar} y_i+d\frac{\eta}{\hbar}\epsilon_{ij}y_j
\end{eqnarray}
where $a, b, c$ and $d$ are dimensionless constants. Now consistency with algebra (\ref{4}) demands 
\begin{eqnarray}
a+d\frac{\eta\theta}{\hbar^2}=1, \  \  \ b+ic\frac{\eta\theta}{\hbar^2}=0, \  \  \ iad+2bc+i\frac{\theta\eta}{\hbar^2}(c^2+d^2)=i.
\end{eqnarray}
Since three equations are not sufficient to fix all the parameters, we find the solutions in terms of $a$ to obtain the following representation of the phase space variables
\begin{eqnarray}
y_i&\rightarrow& y_i\nonumber\\
q_i&\rightarrow& -ia\hbar\frac{\partial}{\partial y_i}\mp i\hbar\sqrt{1-a^2-\frac{\eta\theta}{\hbar^2}}\epsilon_{ij}\frac{\partial}{\partial y_j}
\label{10}\\
&&\pm\frac{\sqrt{1-a^2-\frac{\eta\theta}{\hbar^2}}}{\theta}\hbar y_i+\frac{1-a}{\theta}\epsilon_{ij}\hbar y_j\nonumber
\end{eqnarray}
This representation should have a smooth commutative limit when $(\theta,\eta)\rightarrow 0$. The natural choice $a=1$ does not satisfy this condition. On the other hand if we take $a=\sqrt{1-\frac{\eta\theta}{\hbar^2}}$ then the representation,
\begin{eqnarray}
q_i&\rightarrow& -i\hbar\sqrt{1-\frac{\eta\theta}{\hbar^2}}\frac{\partial}{\partial y_i}+\frac{1-\sqrt{1-\frac{\eta\theta}{\hbar^2}}}{\theta}\hbar\epsilon_{ij}y_j
\label{re}
\end{eqnarray}
has a smooth limit, which is
\begin{eqnarray}
\lim_{\eta\rightarrow 0}\lim_{\theta\rightarrow 0} q_i=-i\hbar\frac{\partial}{\partial y_i}=\lim_{\theta\rightarrow 0}\lim_{\eta\rightarrow 0} q_i
\end{eqnarray}

Noting that the algebra (\ref{4}) is invariant under the transformation $(y,q,\theta,\eta)\rightarrow (q,-y,\eta,\theta)$, we can make this transformation in (\ref{re}) to get the momentum space representation.

Now we set the gravitational well problem on this NC plane, where the vertical direction ($y_1$) is taken to be the Earth's gravitational field and the horizontal direction ($y_2$) is defined by the initial velocity of the particle. The Hamiltonian of this problem is
\begin{eqnarray}
H&=&\frac{1}{2m}(q_1^2+q_2^2)+mgy_1
\nonumber
\end{eqnarray}
where $m$ is the mass of neutron and $g$ is the gravitational acceleration near the surface of the Earth. Using the representation (\ref{re}), this Hamiltonian can be written in the form
\begin{eqnarray}
H&=&\frac{\hbar^2}{2m}\left[-(1-\theta\eta)\left(\frac{\partial^2}{\partial y_1^2}+\frac{\partial^2}{\partial y_2^2}\right)+\left(\frac{1-\sqrt{1-\frac{\theta\eta}{\hbar^2}}}{\theta}\right)^2(y_1^2+y_2^2)\right]\nonumber\\
&&+\frac{\hbar^2}{2m}\left[2i(\sqrt{1-\frac{\theta\eta}{\hbar^2}})\frac{1-\sqrt{1-\frac{\theta\eta}{\hbar^2}}}{\theta}\left(y_1\frac{\partial}{\partial y_2}-y_2\frac{\partial}{\partial y_1}\right)\right]+mgy_1\nonumber
\end{eqnarray}
Making use of the formula $\left(1-x\right)^{1/2}=1-\frac{1}{2}x$ for small (compared to unity) $x$, we simplify the Hamiltonian to get
\begin{eqnarray}
H&=&\frac{1}{2m}\left[-(1-\theta\eta)\hbar^2\left(\frac{\partial^2}{\partial y_1^2}+\frac{\partial^2}{\partial y_2^2}\right)+\left(\frac{\eta}{2\hbar}\right)^2(y_1^2+y_2^2)\right]\nonumber\\
&&+\frac{1}{2m}\left[2i(1-\frac{\theta\eta}{2\hbar^2})\frac{\eta}{2}\left(y_1\frac{\partial}{\partial y_2}-y_2\frac{\partial}{\partial y_1}\right)\right]+mgy_1
\end{eqnarray}
keeping terms only upto first order in the noncommutative parameters, this further reduces to
\begin{eqnarray}
H&=&\frac{1}{2m}\left[-\hbar^2\left(\frac{\partial^2}{\partial y_1^2}+\frac{\partial^2}{\partial y_2^2}\right)+i\eta\left(y_1\frac{\partial}{\partial y_2}-y_2\frac{\partial}{\partial y_1}\right)\right]+mgy_1.
\end{eqnarray}
Since $\theta$ does not appear in the leading order expression of the Hamiltonian{\footnote {This can be confirmed by a different method (general phase space transformation) as discussed in \cite{saurav}.}}, we drop it from the algebra (\ref{4}). Renaming $y$ as $x$ and $-i\hbar\frac{\partial}{\partial y}$ as $p$ we note that $x$ and $p$ are nothing but the canonical pairs of ordinary quantum mechanics, satisfying the algebra
\begin{eqnarray}
[x_i,x_j]=0=[p_i,p_j], \  \  \ [x_i,p_j]=i\hbar\delta_{ij}.
\label{1}
\end{eqnarray}
Thus we write the Hamiltonian in terms of the commutative variables as
\begin{eqnarray}
H&=&H_0-\frac{\eta}{2m\hbar}(x_1p_2-x_2p_1)
\label{H}
\end{eqnarray}
where $H_0$ is the commutative Hamiltonian of the problem
\begin{eqnarray}
H_0=\frac{1}{2m}(p_1^2+p_2^2)+mgx_1.
\label{H0}
\end{eqnarray}
The exact solution of the wave function and the energy eigen values of $H_0$ are given in \cite{lan}. Here we shall treat the extra term of the Hamiltonian as the perturbation and shall take the solutions of the unperturbed Hamiltonian $H_0$ to be the energy levels coming from the WKB approximation,
\begin{eqnarray}
E_n=\left(\frac{9m}{8}[\pi\hbar g(n-\frac{1}{4})]^2\right)^{\frac{1}{3}}=\alpha_{n}g^{\frac{2}{3}} \ ; \ n=1, \ 2, \ 3...
\label{energy}
\end{eqnarray}
We take the following values of different constants 
\begin{eqnarray}
&&\hbar=10.59\times 10^{-35} \ {\textrm {Js}}, \  \  \ g=9.81 \ {\textrm {ms}}^{-2}, \  \  \ m=167.32\times 10^{-29} \ {\textrm {Kg}}\nonumber
\end{eqnarray}
to calculate the first two energy levels of the unperturbed Hamiltonian, which comes out to be:
\begin{eqnarray}
&&E_1=1.392 \ {\textrm {peV}}=2.23\times 10^{-31}{\textrm{J}}, \  \  \ E_2=2.447 \ {\textrm {peV}}=3.92\times 10^{-31}{\textrm{J}}.\nonumber
\end{eqnarray}
Now coming back to the perturbation term of (\ref{H}) we note that, the expectation value $<p_2>=0$. The physical reason is: for a bound state system, the average current flow in a particular direction should be zero. The mathematical derivation of this result is given in \cite{saurav,ba}. Thus we write the complete Hamiltonian as 
\begin{eqnarray}
H&=&\frac{1}{2m}(p_1^2+p_2^2)+mgx_1-\frac{\eta}{2m\hbar}p_2x_1\nonumber\\
&=&\frac{1}{2m}(p_1^2+p_2^2)+mg'x_1
\label{H1}
\end{eqnarray}
This form of the Hamiltonian is quite similar to the unperturbed Hamiltonian (\ref{H0}) . Replacing $g$ by $g'$ in (\ref{energy}) we get the corrected energy values of the Hamiltonian (\ref{H1}) as:
\begin{eqnarray}
E_n+\Delta E_n&=&\alpha_n(g')^{\frac{2}{3}}=\alpha_n(g-\frac{\eta}{2m^2\hbar}<p_2>)^{\frac{2}{3}}\nonumber
\end{eqnarray}
keeping the leading $\eta$-order term we get
\begin{eqnarray}
\Delta E_n=-\frac{\eta}{3gm^2\hbar}<p_2>E_n.\nonumber
\end{eqnarray}
Putting the experimental value $<p_2>=10.91\times 10^{-27} \ {\textrm {Kg m s}}^{-1}$ \cite{nes}we get the following relation between error bars and the noncommutative parameter
\begin{eqnarray}
|\Delta E_1|=2.79\times 10^{29}\eta \ ({\textrm {J}}), \  \  \ |\Delta E_2|=4.90\times 10^{29}\eta \ ({\textrm {J}}).
\label{14}
\end{eqnarray}
Error bars for the above mentioned energy levels are\cite{nes,saurav,ba}
\begin{eqnarray}
&&\Delta E_1^{\textrm {exp}}=6.55\times 10^{-32} \ {\textrm{J}}, \  \  \ \Delta E_2^{\textrm {exp}}=8.68\times 10^{-32} \ {\textrm{J}}.
\end{eqnarray}

Using eq. (\ref{14}), upper bounds for the lowest two energy levels are
\begin{eqnarray}
&&|\eta|\lesssim 2.35\times 10^{-61} \ {\textrm {kg}}^2{\textrm {m}}^2{\textrm {s}}^{-2} \  \ (n=1)\nonumber\\&&|\eta|\lesssim 1.77\times 10^{-61} \ {\textrm {kg}}^2{\textrm {m}}^2{\textrm {s}}^{-2} \  \ (n=2)\nonumber
\end{eqnarray}
These results are in excellent agreement with the numerical results of \cite{ba}. Thus in this article, we have obtained an upper bound on the $\eta$- noncommutativity parameter appearing in the algebra of momenta.

\end{document}